\documentstyle[epsf,12pt]{article}
\begin{document}

\begin{titlepage}
\begin{flushright}
OSU-HEP-01-11\\
BA-02-01\\
January 2002\\
\end{flushright}
\vskip 2cm
\begin{center}
{\Large\bf Eliminating the d = 5 Proton Decay Operators}\\
\vskip 0.4cm
{\Large\bf from SUSY GUTs}
\vskip 1.4cm
{\large\bf
K.S.\ Babu$\,{}^1$ and S.M. Barr$\,{}^{2}$} \\
\vskip 0.5cm
{\it ${}^1\,$Department of Physics, Oklahoma State University\\
Stillwater, OK~~74078, USA\\ [0.1truecm]
${}^2\,$Bartol Research Institute, University of Delaware\\
Newark, DE~~ 19716, USA\\[0.1truecm]
}

\end{center}
\vskip .5cm

\begin{abstract}

A general analysis is made of the question whether the $d=5$
proton decay operators coming from exchange of colored Higgsinos can be
completely eliminated in a natural way in supersymmetric grand unified models.
It is shown that they can indeed be in $SO(10)$ while at the same time naturally
solving the doublet-triplet splitting problem, having only two light Higgs
doublets, and using no more than a single adjoint Higgs field. Accomplishing
all of this requires that the vacuum expectation value of the
adjoint Higgs field be proportional to the generator $I_{3R}$ rather than to
$B-L$, as is usually assumed. It is shown that such models can give
realistic quark and lepton masses.  We also point out a new mechanism
for solving the $\mu$ problem in the context of $SO(10)$ SUSY GUTs.  

\end{abstract}

\end{titlepage}

\section{Introduction}

A major difficulty facing supersymmetric grand unified theories
(SUSY GUTs) \cite{georgi,susyguts} is
the problem of Higgsino-mediated proton decay \cite{hmpd}.
The two Higgs-doublet
superfields of the MSSM, $H_u$ and $H_d$, have in all grand unified models
color-triplet partners, which we may call $H_{uc}$ and $H_{dc}$.
The scalar components of these color-triplets must have masses of order
the GUT scale to avoid catastrophically rapid proton decay. Typically, these
superlarge masses arise from a term in the superpotential of the form
$M_3 H_{uc} H_{dc}$, where $M_3 \sim M_{GUT}$. Given such a term, however,
the fermionic components of $H_{uc}$ and $H_{dc}$ can mediate proton
decay through the diagram shown in Fig. 1. Theoretical
calculations \cite{pdkth1,pdkth2} show that
this diagram typically leads to a proton-decay
rate that is several orders of magnitude
greater than the current limits from super-Kamiokande \cite{superk}. It is
difficult to imagine that
the grand unification framework is totally wrong,
since it explains beautifully the observed
multiplet structure of fermions and the unification of gauge couplings
in a supersymmetric context.  Something must then be greatly suppressing these
dangerous $d=5$ proton-decay operators in SUSY GUTs.  It is this question
that we take up in this paper.

It is apparent that the problem of Higgsino-mediated proton decay is
intimately connected with the well-known ``doublet-triplet splitting problem"
of grand unification, which
is the problem of explaining in a technically natural way how
$H_{uc}$ and $H_{dc}$ acquire GUT-scale masses while leaving their
weak-doublet partners $H_u$ and $H_d$ with only weak-scale masses.
As we shall see, there is a way of suppressing the
dangerous proton-decay operators coming from the diagram of Fig. 1
that is quite simple but that leads to the existence of four
light Higgs doublets rather than the two of the MSSM. Four light doublets
would be incompatible with the unification of gauge couplings unless the
effect of the extra pair of light doublets on the running of the couplings
were accidentally cancelled by the contributions of some other fields.
Such a cancellation would be artificial and unnatural, and so we assume
in this paper that at low energy only the fields of the MSSM exist,
and in particular that there are just two light doublets, $H_u$ and $H_d$.

The question then arises how to suppress the dangerous proton decay
operators while having only two light doublet Higgs.
Various interesting ideas for achieving some degree of suppression
have been proposed in the literature \cite{bb93,suppression}. In this paper,
however, we
are interested in completely killing the dangerous $O(1/M_{GUT})$
operators. That is, we want to find models in which the diagram in Fig. 1
simply does not exist. There is a simple way to do this
in $SO(10)$ models \cite{so10}, as pointed out in \cite{bb93}.
However, this way requires the existence of two
or more Higgs fields in the adjoint representation of $SO(10)$.
Experience has shown that requiring the existence of several adjoints
in a GUT complicates the task of finding realistic string vacua \cite{dienes}. Moreover,
models with multiple adjoint fields would necessarily have a much
more complicated
Higgs sector than models with a single adjoint, as well as
larger GUT-scale threshold corrections to the gauge
couplings. For these reasons it seems desirable to look for models with
only one adjoint field.

The goal we set ourselves in this paper, therefore, is to construct SUSY GUT
models satisfying the following four criteria:
\begin{enumerate}
\item No fine-tuning or
artificial cancellations;
\item Complete suppression of the dangerous
Higgsino-mediated proton-decay amplitude;
\item Unification of gauge
couplings, and therefore only two light Higgs doublets and no light exotic
fields; and
\item An economy of
large representations, and in particular no more than a single field in
the adjoint representation.
\end{enumerate}
As we shall see, $SO(10)$ models which
simultaneously satisfy all of these constraints can indeed
be constructed. The Higgs content of the simplest such models are almost
minimal in a sense that will be explained. To construct such models it
is necessary that the Dimopoulos-Wilczek mechanism \cite{dw}
be implemented in a way
different from that usually assumed. Instead of the expectation value of the
adjoint Higgs field pointing in the $B-L$ direction, it must point in the
$I_{3R}$ direction. We shall show that completely realistic models can
result from this pattern of symmetry breaking, and that these models are
as simple as those based on the usually assumed pattern.

The paper will be organized as follows. In Section 2, we shall review the
problem of doublet-triplet splitting and Higgsino-mediated proton decay
in $SO(10)$. We shall see how the suppression of proton decay, natural
doublet-triplet splitting, and the number of light doublets are related
to each other. We shall see how the existence of two adjoints would allow
a simple resolution of the problems.

In section 3, we construct $SO(10)$ models that resolve the problems but have
only one adjoint, and therefore satisfy all four of our stated criteria.
In part A of section 3, we review some facts about
$SO(10)$ group theory and symmetry-breaking in $SO(10)$ models that
will be needed for the subsequent discussions. In part B of section 3,
we look at $SO(10)$ models having a single adjoint that points in the
$B-L$ direction, and show that all four criteria cannot be satisfied at once:
either the bad proton decay is unsuppressed or there are too many
light doublets. In part C of section 3, we show how models satisfying
all four criteria can be constructed if the VEV of the single adjoint Higgs
points in the $I_{3R}$ direction. We present two examples having nearly
minimal Higgs content. In part D of section 3, we show how realistic quark
and lepton masses can be obtained in simple ways in models of this type.
We also find that these models yield a simple solution to the $\mu$ problem.

In section 4, we abandon the fourth criterion, and study the case of $SO(10)$
with two adjoints. We show how realistic symmetry breaking can be achieved
economically in this case. In section 5, we briefly discuss other groups.
Section 6 has our conclusions.
In Appendix A we describe a third example with a single $I_{3R}$ adjoint
Higgs that suppresses $d=5$ proton decay operators completely.  Appendix B proves that
complete suppression of $d=5$ proton decay operators requires that the Higgs
doublet should arise, at least partly, from vector representation of $SO(10)$.

\section{A review of the Higgsino-mediated proton decay problem}

First consider the case of $SU(5)$ \cite{su5}.
The Higgs of the MSSM are
contained in the representations $\overline{{\bf 5}}_H + {\bf 5}_H$.
For brevity we shall use the following notation for the doublet and triplet
representations of the standard model: ${\bf 2} \equiv (1, 2, \frac{1}{2})$,
$\overline{{\bf 2}} \equiv (1,2, -\frac{1}{2})$, ${\bf 3} \equiv
(3, 1, -\frac{1}{3})$, $\overline{{\bf 3}} \equiv (\overline{3},
1, \frac{1}{3})$. Thus, the MSSM
Higgs multiplets are $H_u = {\bf 2}$ and $H_d = \overline{{\bf 2}}$, and the
$SU(5)$ fundamental representations decompose as ${\bf 5} = {\bf 2} + {\bf 3}$
and $\overline{{\bf 5}} = \overline{{\bf 2}} + \overline{{\bf 3}}$. In minimal
$SU(5)$ the doublet-triplet splitting is done by a combination of
an adjoint Higgs, ${\bf 24}_H$, and a singlet Higgs, ${\bf 1}_H$ (or equivalently
a bare mass term).
The relevant terms in the superpotential are $\overline{{\bf 5}}_H
(\lambda {\bf 24}_H + \lambda' {\bf 1}_H) {\bf 5}_H$. The VEV of the
adjoint must point in the hypercharge direction: $\langle {\bf 24}_H
\rangle \propto
{\rm diag} ( -\frac{1}{2}, - \frac{1}{2}, \frac{1}{3}, \frac{1}{3},
\frac{1}{3})$, whereas the VEV of the singlet is obviously proportional
to the identity: $\langle {\bf 1}_H \rangle \propto {\rm diag} (1,1,1,1,1)$.
If the couplings $\lambda$ and $\lambda'$ are fine-tuned to be in the
right ratio, then the effective mass term can be of the form
${\rm diag} (0,0, \frac{5}{6}, \frac{5}{6},\frac{5}{6})$. This will give mass
to the $\overline{{\bf 3}}-{\bf 3}$ pair, but not to the
$\overline{{\bf 2}}-{\bf 2}$ pair.
Schematically, we will denote such a mass term for triplets but not
doublets by

\begin{equation}
\left( \begin{array}{c} \overline{2} \\ \overline{3} \end{array} \right)
\begin{array}{c} \\ \longleftrightarrow \end{array}
\left( \begin{array}{c} 2 \\ 3 \end{array} \right)~.
\end{equation}

\noindent
We will call such a mass a ``3-mass."
This minimal $SU(5)$ scheme satisfies two of our criteria:
it uses only a single adjoint Higgs and has the correct number of light
doublets. However, it requires a severe fine-tuning to make the Higgs doublets
light, and it does not suppress the bad proton decay. The Higgsino-mediated
proton decay is unsuppressed because there is a mass term connecting
the $\overline{3} = H_{dc}$ and $3 = H_{uc}$ of Higgs, allowing the diagram
of Fig. 1 to be drawn.

The situation is much better in minimal $SO(10)$. In minimal SUSY $SO(10)$
the two doublets of the MSSM are contained in a single vector representation,
which we shall call, for reasons that will be clear shortly, ${\bf 10}_{1H}$.
Decomposing under the standard Georgi-Glashow $SU(5)$ subgroup of
$SO(10)$, one has ${\bf 10}_{1H} = \overline{{\bf 5}}_{1H} +
{\bf 5}_{1H} = (\overline{{\bf 2}}_{1H} + \overline{{\bf 3}}_{1H})
+ ({\bf 2}_{1H} + {\bf 3}_{1H})$. The doublet-triplet splitting can be
achieved by
coupling this vector to an adjoint Higgs, ${\bf 45}_H$. Since the adjoint
of $SO(10)$ in the fundamental representation is a 10-by10 antisymmetric
matrix, two distinct vector Higgs must appear
in this coupling: ${\bf 10}_{1H} {\bf 45}_H {\bf 10}_{2H}$. A great advantage
of $SO(10)$ over $SU(5)$ is that the VEV of the ${\bf 45}$, when written
in the fundamental representation, can take the form
$\langle {\bf 45}_H \rangle \propto {\rm diag}(a_1, a_2, a_3, a_4, a_5)
\otimes i \tau_2$, where in contrast to $SU(5)$ there is no requirement that
the trace $\Sigma_i a_i$ vanish. Thus one can have
$\langle {\bf 45}_H \rangle \propto {\rm diag}(0,0,1,1,1)
\otimes i \tau_2$. In fact, this form is just proportional to the
$SO(10)$ generator $B-L$. Such a VEV will give mass to the triplets in
${\bf 10}_{1H}$ and ${\bf 10}_{2H}$ while leaving the doublets massless,
i.e.give a 3-mass.
This is the so-called Dimopoulos-Wilczek mechanism \cite{dw},
and we shall call the
adjoint VEV $\langle {\bf 45}_H \rangle \propto B-L$ the
``Dimopoulos-Wilczek form." Schematically,
the masses that result from the term
${\bf 10}_{1H} {\bf 45}_H {\bf 10}_{2H}$ can be denoted

\begin{equation}
\left( \begin{array}{c} \overline{2}_1 \\ \overline{3}_1 \end{array} \right)
\begin{array}{c} \\ \longleftrightarrow \end{array}
\left( \begin{array}{c} 2_2 \\ 3_2 \end{array} \right) \;\;\;\;\;\;
\left( \begin{array}{c} \overline{2}_2 \\ \overline{3}_2 \end{array} \right)
\begin{array}{c} \\ \longleftrightarrow \end{array}
\left( \begin{array}{c} 2_1 \\ 3_1 \end{array} \right).
\end{equation}

If we say that ${\bf 10}_{1H}$ couples to the quarks and leptons, but that
${\bf 10}_{2H}$ does not, then the color triplets that couple to the quarks
and leptons are the $\overline{3}_1$ and $3_1$. One sees from Eq. (2), however,
that there is no mass term connecting $\overline{3}_1$ and $3_1$
to each other, directly or indirectly. Thus the diagram of Fig. 1 cannot be
drawn, as the mass insertion denoted there by $M_3$ vanishes. This
implies the absence of the dangerous Higgsino-mediated proton decay operator.

While the term ${\bf 10}_{1H} {\bf 45}_H {\bf 10}_{2H}$ allows the
doublet-triplet splitting to be achieved without the kind of fine-tuning
necessary in minimal $SU(5)$, and also allows a suppression of the dangerous
proton decay, one sees immediately from Eq. (2) that there is a serious
problem: there are four light doublets not two, since there is no mass term
for any of the doublets in ${\bf 10}_{1H}$ or ${\bf 10}_{2H}$. This would
destroy the unification of gauge couplings. What is needed is to give mass
to the doublets $\overline{2}_2$ and $2_2$. The obvious way to do this is
to posit the existence of an explicit mass term $M {\bf 10}_{2H}{\bf 10}_{2H}$.
(A term that, like this, gives mass to both triplets and doublets we will
call a ``5-mass.") However, this would lead to the situation shown in Eq. (3):

\begin{equation}
\left( \begin{array}{c} \overline{2}_1 \\ \overline{3}_1 \end{array} \right)
\begin{array}{c} \\ \longleftrightarrow \end{array}
\left( \begin{array}{c} 2_2 \\ 3_2 \end{array} \right)
\begin{array}{c} \longleftrightarrow \\ \longleftrightarrow \end{array}
\left( \begin{array}{c} \overline{2}_2 \\ \overline{3}_2 \end{array} \right)
\begin{array}{c} \\ \longleftrightarrow \end{array}
\left( \begin{array}{c} 2_1 \\ 3_1 \end{array} \right).
\end{equation}

\noindent
Now, indeed, there is the correct number of light doublets, but the dangerous
proton-decay has returned, since there is an effective mass term linking
$\overline{3}_1$ and $3_1$ after the triplets $\overline{3}_2$ and $3_2$
have been integrated out, as shown in Fig. 2.

These simple examples
illustrate the close link between the problem of suppressing Higgsino-mediated
proton decay, and the problem of obtaining the correct number of light
doublets. There is a way out of the dilemma, as was pointed out in \cite{bb93}.
Imagine that a second adjoint Higgs ${\bf 45}'_H$ exists, whose VEV
lies in the $I_{3R}$ direction:
$\langle {\bf 45}'_H \rangle \propto {\rm diag}(1,1,0,0,0) \otimes i \tau_2$.
If this couples by a term of the form ${\bf 10}_{2H} ({\bf 45}'_H)^2
{\bf 10}_{2H}/M$, then it will give a mass to the
doublets in the ${\bf 10}_{2H}$ but not the triplets. Such a term we call a
``2-mass." The result can be
written as:

\begin{equation}
\left( \begin{array}{c} \overline{2}_1 \\ \overline{3}_1 \end{array} \right)
\begin{array}{c} \\ \longleftrightarrow \end{array}
\left( \begin{array}{c} 2_2 \\ 3_2 \end{array} \right)
\begin{array}{c} \longleftrightarrow \\ {\rm } \end{array}
\left( \begin{array}{c} \overline{2}_2 \\ \overline{3}_2 \end{array} \right)
\begin{array}{c} \\ \longleftrightarrow \end{array}
\left( \begin{array}{c} 2_1 \\ 3_1 \end{array} \right).
\end{equation}

\noindent
Here one sees that the dangerous proton decay has been killed
and that there are
only two light doublets. The great price that has been paid for this
success is that the model has two adjoint Higgs. As will be shown in
section 4, satisfactory SUSY $SO(10)$ models can be constructed that have
two adjoints with the desired VEVs. However, as noted earlier, there are
theoretical drawbacks to models with multiple adjoint Higgs.
The question therefore arises whether the dangerous
proton decay can be suppressed in SUSY $SO(10)$ with only a single adjoint
field, without fine tuning and without extra light Higgs doublets.
This is the issue that we address in the next section.

\section{$SO(10)$ with a single adjoint Higgs field}

\noindent
{\bf A. Review of symmetry breaking in $SO(10)$}

\vspace{0.2cm}

\noindent
{\it (i) Breaking down to the Standard Model.}
To break $SO(10)$ down to the Standard Model group $SU(3) \times SU(2) \times
U(1)$ two kinds of Higgs fields are required \cite{bb93, bb}. First, there
must be a sector of
Higgs that give the right-handed neutrinos a mass and break the rank of the
group from five to four. This will leave unbroken an $SU(5)$ subgroup.
This sector must include, at least, either a $\overline{{\bf 126}}_H$ or
a $\overline{{\bf 16}}_H$. These are generally assumed, in order not
to break SUSY at the GUT scale, to be accompanied by fields in their conjugate
representations. There are some
difficulties associated with having $\overline{{\bf 126}}_H + {\bf 126}_H$;
in particular obtaining them from string theory \cite{dienes}, and avoiding the unified
coupling becoming non-perturbative below the Planck scale.
We will assume, therefore, that the breaking of the rank of the group is done
by $\overline{{\bf 16}}_H + {\bf 16}_H$, and will call the
sector containing these fields the ``spinor Higgs sector".
Second, there must be a sector of Higgs that complete the
breaking down to $SU(3) \times SU(2) \times U(1)$ and do the doublet-triplet
splitting. This sector must include at least one adjoint, and we will
therefore call it the ``adjoint Higgs sector".

The GUT-scale VEVs of the $\overline{{\bf 16}}_H + {\bf 16}_H$ must lie in the
``right-handed neutrino direction". That is, under the Standard Model
group the components that get VEVs are in $(1,1,0)$. As for the
adjoint Higgs sector, in order to do doublet-triplet splitting there
must be an adjoint that has VEV in either the $B-L$ direction or the
$I_{3R}$ direction. (Recall that $SO(10) \supset SU(4)_c \times
SU(2)_L \times SU(2)_R$, and that $B-L$ is a generator of $SU(4)_c$
while $I_{3R}$ is a generator of $SU(2)_R$.)

\vspace{0.2cm}

\noindent
{\it (ii) Doublet-triplet splitting.}
The generator $B-L$ in
the fundamental (i.e. vector) representation is the $10 \times 10$
antisymmetric matrix $-\frac{2}{3} {\rm diag}(0,0,1,1,1) \times i \tau_2$.
This can be seen from the fact that the Higgs doublets in an $SO(10)$ vector
have $B-L =0$,
whereas their color-triplet partners have $B-L = \pm \frac{2}{3}$.
We already saw in the previous section how having an adjoint VEV in the $B-L$
direction (the ``Dimopoulos-Wilczek form")
allows a natural solution to the doublet-triplet splitting
problem: by coupling vector Higgs to such an adjoint through a term of the form
${\bf 10}_{1H} {\bf 45}_H {\bf 10}_{2H}$, the doublets remain massless
since they have $B-L = 0$, whereas the triplets get mass.

It is less obvious how having an adjoint with VEV in the $I_{3R}$
direction would allow a natural solution of the doublet-triplet splitting
problem. The generator $I_{3R}$ in the vector representation is the
$10 \times 10$ antisymmetric matrix $\frac{1}{2} {\rm diag} (1,1,0,0,0)
\times i \tau_2$.
We will therefore call
$\langle {\bf 45}_H \rangle \propto I_{3R}$ the ``upside-down
Dimopoulos-Wilczek form". That $I_{3R}$ looks like this
is clear from the fact that the MSSM Higgs doublets
are in a $(1,2,2)$ of $SU(4)_c \times SU(2)_L \times SU(2)_R$, while
their color-triplet partners are in a $(6,1,1)$ (together making up a
${\bf 10}$ of $SO(10)$). If an adjoint with VEV in the $I_{3R}$ direction
were to couple to Higgs in the vector representation, therefore, it
would make the doublets heavy and leave the color triplets massless, i.e.
give a 2-mass ---
precisely the opposite of what is needed for the doublet-triplet splitting.

However, there are also components in the {\it spinors} of $SO(10)$ that
have the right quantum numbers to be the MSSM Higgs doublets.
Under the $SU(5)$ subgroup of $SO(10)$,
the spinors decompose as ${\bf 16} \longrightarrow
{\bf 10} + \overline{{\bf 5}} + {\bf 1}$ and $\overline{{\bf 16}}
\longrightarrow \overline{{\bf 10}} + {\bf 5} + {\bf 1}$.
From the point of view of the Standard Model the
${\bf 5}_{\overline{16}}$ and $\overline{{\bf 5}}_{16}$
have the same quantum numbers as the ${\bf 5}_{10}$ and
$\overline{{\bf 5}}_{10}$. That is, they contain just the same kinds
of doublets and triplets. However, $I_{3R}$ acts differently on them.
The doublets (i.e. the $(1,2, \pm \frac{1}{2})$) in the $SO(10)$
spinors have the same quantum numbers as the lepton doublets
$L = ( \nu, \ell^-)$ or their conjugates, and are therefore singlets
under $SU(2)_R$ and have $I_{3R} = 0$. The triplets (i.e.
$(3, 1, - \frac{1}{3}) + {\rm conj.}$) in the $SO(10)$ spinors have
the same quantum numbers as the $d_R$ or their conjugates, and are
therefore doublets under $SU(2)_R$ and have $I_{3R} = \pm \frac{1}{2} \neq 0$.
Consequently, if one couples $\overline{{\bf 16}}_H {\bf 45}_H
{\bf 16}_H$, where $\langle {\bf 45}_H \rangle \propto I_{3R}$, then
the doublets remain massless and their triplet partners get mass
(i.e. there is a 3-mass), just
as needed for the doublet-triplet splitting \cite{dvali, balphas}.

Let us reiterate.
An adjoint in the $B-L$ direction, if it couples to {\it vectors},
gives a 3-mass. On the other hand,
an adjoint in the $I_{3R}$ direction, if it couples to {\it vectors},
gives a 2-mass, whereas if it couples to {\it spinors},
gives a 3-mass. It is the fact that an adjoint in the $I_{3R}$ direction can
play this double role, giving either 2-masses or 3-masses,
depending on what it couples to,
that will be critically important when we discuss proton decay later in
this section.

It should be noted that the spinor Higgs we are talking about here
would be ones that contain doublets that are involved
in the breaking of $SU(2) \times U(1)$ at the weak scale. They would
not be the same spinor Higgs, referred to above, needed to break
$SO(10)$ down to the Standard Model group at the GUT scale. In
order to distinguish these kinds of
spinor from each other, we will from now on
denote those spinors that have GUT-scale VEVs in the $(1,1,0)$ direction
{\it without} a prime, and denote those spinors that have no GUT-scale VEVs,
but possibly have weak-scale VEVs in
the $(1,2, \pm \frac{1}{2})$ direction, with
one or more primes.

\vspace{0.2cm}

\noindent
{\it (iii) Avoiding colored goldstone bosons.}
Returning now to the breaking of $SO(10)$ to the Standard Model group
at the GUT scale,
we recall that two sectors are needed, an adjoint Higgs sector and a
spinor Higgs sector.
It is critical that these two Higgs sectors be coupled together, directly
or indirectly. If they were not, then uneaten colored Goldstone bosons would
result. The point is that there are generators in $SO(10)/(SU(3) \times
SU(2) \times U(1))$ that are broken by both sectors. If the two sectors
were not coupled together in the superpotential, then both sectors would have
a goldstone boson corresponding to each of these broken generators. However,
only one goldstone for each broken generator gets eaten by the Higgs
mechanism; thus there would be uneaten goldstones left over. It is simple
to check that these uneaten goldstones would include ones in
$(3,2, \frac{1}{6}) +
(\overline{3}, 2, - \frac{1}{6})$. These would only get mass when
supersymmetry breaks, and would therefore have a ruinous effect
on the running of the gauge couplings.

The coupling of the spinor Higgs sector to the adjoint Higgs sector is
a non-trivial problem, for the simplest ways to couple them together
destabilize the form of the adjoint vacuum expectation value needed
for doublet-triplet splitting, i.e. the Dimopoulos-Wilczek form or
alternatively the upside-down Dimopoulos-Wilczek form.
For example, consider the coupling
$\overline{{\bf 16}}_H {\bf 45}_H {\bf 16}_H$, where the spinors
are those that have GUT-scale VEVs in the $SU(5)$-singlet direction.
Then the $F_{{\bf 45}_{H}} = - (\partial W/\partial ({\bf 45}_H))^*$
will contain
a term $-(\overline{{\bf 16}}_H {\bf 16}_H)^*$ that drives ${\bf 45}_H$
away from the $B-L$ and $I_{3R}$ directions. One can see this also by
considering the $F$ terms for the spinors: $\partial W/\partial
(\overline{{\bf 16}}_H) \supset {\bf 45}_H {\bf 16}_H$. If the adjoint
points in the $B-L$ or $I_{3R}$ directions then this will not vanish,
as typically required to avoid breaking SUSY at the GUT scale.

The simplest way to couple the spinor Higgs sector to the adjoint
Higgs sector without destabilizing the adjoint VEV was proposed in
\cite{barrraby}. The relevant terms are of the form

\begin{equation}
\overline{{\bf 16}}_H ({\bf 45}_H + {\bf 1}_H) {\bf 16}'_H +
\overline{{\bf 16}}'_H ({\bf 45}_H + {\bf 1}_H) {\bf 16}_H.
\end{equation}

\noindent
Here, in accordance with the notational convention explained above,
the unprimed spinors have GUT-scale VEVs that break $SO(10)$ to
$SU(5)$, whereas the primed spinors have no GUT-scale VEV, though they
may get a weak-scale VEV when SUSY breaks. In the supersymmetric limit,
therefore, we can set the VEVs of the primed spinors to zero. Then the
$F$ terms corresponding to the fields $\overline{{\bf 16}}_H$, ${\bf 16}_H$,
${\bf 45}_H$, and ${\bf 1}_H$ will vanish, as needed to keep the
adjoint VEV from being destabilized. The $F$ terms corresponding to
the primed spinors must also vanish, and this is the significance of the
singlets appearing in Eq. (5). These are assumed to have no other terms in
the superpotential that would fix their values, so that they are ``sliding
singlets" that are free to adjust their values to make the $F$ terms of
the primed spinors zero.
It is straightforward to show that the terms in Eq. (5) are sufficient to
prevent the appearance of uneaten goldstone bosons, and do not destabilize
the adjoint VEVs.

It was shown in \cite{barrraby} that a satisfactory breaking of
$SO(10)$ down to the Standard Model with a natural doublet-triplet
splitting can be achieved with the following set of Higgs fields:
${\bf 45}_H$, $\overline{{\bf 16}}_H$, ${\bf 16}_H$,
$\overline{{\bf 16}}'_H$, ${\bf 16}'_H$, ${\bf 10}_{1H}$, ${\bf 10}_{2H}$,
and some number of singlets. This is the minimal set of Higgs
that any SUSY $SO(10)$ GUT must have. The adjoint and the unprimed
spinors are needed to break $SO(10)$; the primed spinors are needed
to couple the adjoint to the unprimed spinors as in Eq. (5). The two
vectors are needed for the Dimopoulos-Wilczek doublet-triplet splitting
via the term ${\bf 10}_{1H} {\bf 45}_H {\bf 10}_{2H}$.

We shall see later in this section that the bad proton-decay operators
can be completely suppressed in models with a Higgs sector that is
almost as small as this minimal set.

\vspace{0.2cm}

\noindent
{\bf (B) The case of one adjoint with $\langle {\bf 45}_H \rangle \propto B-L$}

\vspace{0.2cm}

Let us consider SUSY $SO(10)$ models that have only one adjoint, whose VEV
is in the $B-L$ direction, and no representations larger than the
adjoint (such as the ${\bf 210}$).
It is not difficult to prove that such a model
cannot satisfy all four of our criteria. If the dangerous proton decay mediated
by the color-triplet Higgsinos is completely suppressed, then there
must be at least four light Higgs doublets.

The proof begins with the fact that, if there is only one adjoint
and it points in
the $B-L$ direction, and if there are no larger representations
than the adjoint, then
there can be only 5-masses or 3-masses, but not 2-masses. The next step is
to establish the conditions under which massless doublets or triplets
exist.

Suppose there is a set $S$ of $\overline{{ N}}$'s and ${ N}$'s of
Higgs fields which are anti-fundamentals and fundamentals of $SU(N)$
with $n_{\overline{N}}$ of the former and $n_N$ of the latter.
$N$ will be a ${\bf 3}$ of $SU(3)$ or ${\bf 2}$ of $SU(2)$ for us.
Define $\nu = n_N - n_{\overline{N}}$. 
If two multiplets (say $\overline{N}_i$ and $N_j$) are connected by 
a mass term (say $m \overline{N}_i N_j$), then we will say that 
they are ``linked". If every multiplet in the set $S$ is connected to
every other multiplet in $S$ through a series of such links, then we
will say that $S$ is a ``connected set."
Let us also assume that no multiplet in $S$ has a mass term that connects it
with any
multiplet outside $S$. We will then say that $S$ is ``isolated".
It is obvious that any set of $N$'s and $\overline{N}$'s can be uniquely
decomposed into isolated, connected sets.
If one has an isolated, connected set $S$ of $\overline{N}$'s and $N$'s,
and there are no accidental cancellations, then, for $\nu =0$ there will
be no massless multiplets in $S$, for $\nu$ positive there will be
$\nu$ massless $N$'s in $S$, and for $\nu$ negative there will be $|\nu|$
massless $\overline{N}$'s in $S$.

Now consider the ``massless" doublets $H_u$ and $H_d$ of the MSSM
(massless in the sense that they have no GUT-scale mass).
Since $H_u$ is massless, it must belong to an isolated, connected set
of doublets (call it $D_u$) having $n_2 - n_{\overline{2}} \geq 1$.
And since we do not want more than one massless ${\bf 2}$, it must be that
$n_2 - n_{\overline{2}} = 1$ for $D_u$. Similarly, $H_d$ must belong
to an isolated, connected set of doublets (call it $D_d$) that has
$n_2 - n_{\overline{2}} = -1$.

Consider the colored partners of the doublets in $D_u$. These must form
a connected set, $T_u$, since $D_u$ is connected and there are no 2-masses.
However, $T_u$ does not have to be isolated (since there can be 3-masses),
but can be connected to
other triplets to form a larger isolated, connected set, which we will call
$T'_u$. Since we do not want any triplets to be massless, it must be
(if there is no fine-tuning) that $n_3 = n_{\overline{3}} = 0$ for
$T'_u$. Denote the set of doublet partners of $T'_u$ by $D'_u$.
Then $D'_u$ is isolated, because $T'_u$ is and there are no 2-masses.
Moreover, it is
obvious that $n_2 = n_{\overline{2}}$ for $D'_u$, since the analogous
condition holds for $T'_u$. However,
$D'_u \supset D_u$, and $D_u$ has $n_2 - n_{\overline{2}} = 1$. Thus
the complement of $D_u$ in $D'_u$ must be an isolated set having
$n_2 - n_{\overline{2}} = -1$. Thus, $D'_u$ contains both a massless
${\bf 2}$ and a massless $\overline{{\bf 2}}$. This massless
$\overline{{\bf 2}}$ is {\it not} the $H_d$, however. If it were, then
the color-triplet partners $H_{uc}$ and $H_{dc}$ would both be in the
connected set $T'_d$, and the bad proton-decay amplitude would not be
suppressed. Thus there are at least {\it two} massless $\overline{{\bf 2}}$'s.
Exactly analogous reasoning shows that there are at least two massless
${\bf 2}$'s. Altogether, then, there are at least four massless doublets
if the bad proton decay is completely suppressed.

This proves the theorem that in $SO(10)$ with only one adjoint, which
points in the $B-L$ direction, and no representations larger than the adjoint,
the bad color-triplet-Higgsino-mediated proton decay cannot
be completely suppressed unless there are at least four light doublets.
The root of the problem is the fact that there are no ``2-masses."

Now, suppose instead that in some $SO(10)$ model there were 2-masses and
5-masses, but no 3-masses. An even worse disaster would result.
The set we called $T_u$ would be isolated (though not necessarily
connected), since $D_u$ is. Moreover, the set $T_u$ satisfies
$n_3 - n_{\overline{3}} = 1$, since $D_u$ satisfies the analogous relation.
Consequently, there would be a massless ${\bf 3}$. Similarly, there
would be a massless $\overline{{\bf 3}}$. Disastrously rapid proton
decay would result.

One concludes, therefore, that to suppress proton decay and have only two
light doublets without fine-tuning, there must be {\it both}
3-masses {\it and} 2-masses. An example is the situation shown in Eq. (4).
But, we have already seen that with a single adjoint which points in
the $I_{3R}$ direction, one can have both 2-masses and 3-masses.
Thus the possibility exists that all four of our criteria can be satisfied
in this case. In the next part of this section, we will see that this
is in fact so.

\vspace{0.2cm}

\noindent
{\bf (C) The case of one adjoint with $\langle {\bf 45}_H \rangle \propto
I_{3R}$}

\vspace{0.2cm}

\noindent
{\it (i) Ways to get 2-masses and 3-masses.}
We have already seen that an adjoint pointing in the $I_{3R}$ direction can
produce a 3-mass through a term of the form
$\overline{{\bf 16}}'_H \langle {\bf 45}_H \rangle {\bf 16}'_H$.
(It is important that the spinors here have no GUT-scale VEVs, so as not
to destabilize the VEV of the adjoint, as explained in part (A) of this
section. Hence the primes on these multiplets.)
We have also seen that an adjoint pointing in the $I_{3R}$ direction
can produce a 2-mass through a term of the form
${\bf 10}_{1H} \langle {\bf 45}_H \rangle {\bf 10}_{2H}$. But, in fact,
there are two other ways that such an adjoint can lead to 2-masses, as
we will now show.

Consider the terms
$\overline{{\bf 16}}_H(({\bf 45}_H)^2/M + {\bf 1}_H) {\bf 16}'_H$
and $\overline{{\bf 16}}'_H(({\bf 45}_H)^2/M + {\bf 1}_H) {\bf 16}_H$,
with $\langle {\bf 45}_H \rangle =
I_{3R} A$. The $F$ terms of the primed spinors will force
$[(\langle {\bf 45}_H \rangle)^2/M + \langle {\bf 1}_H \rangle]
\langle {\bf 16}_H \rangle = 0$. Since the component of ${\bf 16}_H$
that has the GUT-scale VEV has $I_{3R} = -\frac{1}{2}$, the singlet
VEV must slide to the value $\langle {\bf 1}_H \rangle = -\frac{1}{4} A^2/M$.
Thus, these terms leave massless those components
of the ${\bf 16}$'s having $I_{3R} = \pm \frac{1}{2}$. In particular,
the $(3, 1, -\frac{1}{3}) \equiv {\bf 3}$, $(\overline{3}, 1, \frac{1}{3})
\equiv \overline{{\bf 3}}$,
$(3,1,\frac{2}{3})$, $(\overline{3}, 1, -\frac{2}{3})$, $(1,1,1)$,
$(1,1,-1)$, and $(1,1,0)$ will be left massless, whereas the
$(1,2, \frac{1}{2}) \equiv {\bf 2}$ and $(1,2, -\frac{1}{2}) \equiv
\overline{{\bf 2}}$ will be be made massive.

There is a third way to get 2-masses from $\langle {\bf 45}_H \rangle
\propto I_{3R}$, and that is through a quartic term
$(\langle {\bf 45}_H \rangle \langle {\bf 16}_H \rangle)
{\bf 16}_H {\bf 10}_H$ in which the two spinors are the same field, and
are therefore contracted symmetrically. One way to see this is by
decomposing the representations under $SU(4)_c \times SU(2)_L \times
SU(2)_R$. Under this subgroup, the GUT-scale VEV of the adjoint is in
the $(1,1,3)$ (since it is proportional to $I_{3R}$), the GUT-scale VEV
of the spinor is in the $(\overline{4}, 1,2)$, the color-triplet Higgs
in the ${\bf 10}_H$ are in $(6,1,1)$, and the color-triplet, weak-singlet Higgs
in the ${\bf 16}_H$ is in $(\overline{4},1,2)$. These four multiplets can be
contracted together, but only in the antisymmetric product of the
two spinors, for $(6,1,1) \times (1,1,3) = (6,1,3)$, which is in
$((\overline{4}, 1,2) \times(\overline{4}, 1,2))_{antisym}$.
Consequently, the triplets are not given mass by
this term. It is easily seen that the doublet Higgs do get mass. Thus this
term does give a 2-mass.

Even if the term is $(\langle {\bf 45}_H \rangle \langle {\bf 16}_H \rangle)
{\bf 16}'_H {\bf 10}_H$, where the two spinors are different fields, a
2-mass can result if the term arises from integrating out a vector field.
Consider, for example a model in which there are the following
couplings: ${\bf 16}'_H {\bf 16}_H {\bf 10}'_H$, $M {\bf 10}'_H {\bf 10}'_H$,
and ${\bf 10}'_H {\bf 45}_H {\bf 10}_H$. If one integrates out the
${\bf 10}'_H$, as shown in Fig. 3, then the following effective term results
$[{\bf 16}'_H {\bf 16}_H]_{10} [{\bf 45}_H {\bf 10}_H]_{10}$. The subscripts
on the brackets indicate that the product of fields in the bracket is
contracted into a ${\bf 10}$. That means that even though the two spinors
are distinct fields thay are here contracted symmetrically, and therefore
a 2-mass results, by the same reasoning as before. (One can also see this
more directly from the fact that the term ${\bf 10}'_H {\bf 45}_H {\bf 10}_H$
gives only a 2-mass.)

Now we shall see how these kinds of couplings can be used to construct
simple models which satisfy the four criteria.

\vspace{0.2cm}

\noindent
{\it (ii) A nearly minimal model.}

The following is a nearly minimal model of doublet-triplet splitting
in SUSY $SO(10)$:

\begin{equation}
\begin{array}{ccl}
W_{2/3} & = & \overline{{\bf 16}}_H ({\bf 45}_H + {\bf 1}_H) {\bf 16}'_H
+ \overline{{\bf 16}}'_H ({\bf 45}_H + {\bf 1}_H) {\bf 16}_H \\
& & \\
& & + \overline{{\bf 16}}'_H {\bf 45}_H {\bf 16}'_H +
m \overline{{\bf 16}}^{\prime \prime}_H {\bf 16}^{\prime \prime}_H \\
& & \\
& & + {\bf 16}_H {\bf 16}_H {\bf 10}_H +
\overline{{\bf 16}}_H \overline{{\bf 16}}_H {\bf 10}_H \\
& & \\
& & + [{\bf 16}^{\prime \prime}_H {\bf 16}_H]_{10} [{\bf 45}_H {\bf 10}_H]_{10}.
\end{array}
\end{equation}

\noindent
The coupling constants have been suppressed. The last term is the one
just discussed, that comes from integrating out a vector and gives a
2-mass.

Note that the first two terms in Eq. (6) are just the same as those given
in Eq. (5).
These terms must exist in any case to couple the adjoint Higgs sector to the
spinor Higgs sector in order to avoid goldstone modes, as explained in
Part A (iii) of this section. The only Higgs multiplets appearing in Eq. (6)
that go beyond the minimal set described in Part A (iii) is the pair
of spinors $\overline{{\bf 16}}^{\prime \prime}_H + {\bf 16}^{\prime \prime}_H$.

If one decomposes the Higgs multiplets appearing in Eq. (6)
under the $SU(5)$ subgroup of
$SO(10)$, there are altogether four $\overline{{\bf 5}}$
(in the ${\bf 16}_H$, ${\bf 16}'_H$, ${\bf 16}^{\prime \prime}_H$, and
${\bf 10}_H$) and four ${\bf 5}$ (in the $\overline{{\bf 16}}_H$,
$\overline{{\bf 16}}'_H$, $\overline{{\bf 16}}^{\prime \prime}_H$, and
${\bf 10}_H$). Thus, the doublet (or triplet) mass matrices are four-by-four:

\begin{equation}
(\overline{{\bf 5}}_{16_H}, \overline{{\bf 5}}_{16'_H},
\overline{{\bf 5}}_{16^{\prime \prime}_H}, \overline{{\bf 5}}_{10_H})
\left( \begin{array}{cccc}
0 & \langle(45_H + 1_H)\rangle & 0 & \langle 16_H \rangle \\ \\
\langle(45_H + 1_H)\rangle & \langle 45_H \rangle_3 & 0 & 0 \\ \\
0 & 0 & m & \langle 45_H 16_H \rangle_2 \\ \\
\langle \overline{16}_H \rangle & 0 & 0 & 0 \\ \end{array} \right)
\left( \begin{array}{c}
{\bf 5}_{\overline{16}_H} \\ \\ {\bf 5}_{\overline{16}'_H} \\ \\
{\bf 5}_{\overline{16}^{\prime \prime}_H} \\ \\ {\bf 5}_{10_H} \\
\end{array} \right).
\end{equation}

\noindent
All of the masses here are 5-masses (i.e. masses for both the doublets
and triplets) except for two of them. There is a 2-mass that comes
from the quartic term referred to above, which we indicate by a subscript 2
in Eq. (7), and a 3-mass that comes from $\overline{{\bf 16}}'_H
\langle {\bf 45}_H \rangle {\bf 16}'_H$, which we indicate by a subscript
3.

The triplet mass matrix, which is gotten simply by setting
the 2-mass in Eq. (7) to zero, is of rank four, so that all the triplets
obtain GUT-scale masses as they must. The doublet mass matrix, gotten by
setting the 3-mass to zero in Eq. (7), is of rank three. There is therefore
exactly one massless pair of doublets, as there should be.
The massless $\overline{{\bf 2}}
\equiv H_d$ is a linear combination of the doublets in ${\bf 16}'_H$
and ${\bf 10}_H$. The massless ${\bf 2} \equiv H_u$ is a linear combination
of the doublets in $\overline{{\bf 16}}'_H$,
$\overline{{\bf 16}}^{\prime \prime}_H$, and ${\bf 10}_H$.
If we suppose that only ${\bf 16}'_H$ and
$\overline{{\bf 16}}^{\prime \prime}_H$ couple to quarks and leptons,
then it can be shown that the proton-decay diagram in Fig. 1 cannot be
drawn, and that the dangerous proton-decay amplitude is thus
completely suppressed.

Some of these statements are not completely obvious from a cursory
inspection of the mass matrix in Eq. (7). A way to see them more
clearly is to represent the mass terms in Eq. (7) schematically using
the same notation as in Eqs. (1)-(4).

\begin{equation}
\begin{array}{l}
\;\; \overline{{\bf 5}}_{10_H} \;\;\;\;\;\;\;\;\; {\bf 5}_{\overline{16}_H}
\;\;\;\;\;\;\;\;\; \overline{{\bf 5}}_{16'_H} \;\;\;\;\;\;\;\;\;
{\bf 5}_{\overline{16}'_H} \;\;\;\;\;\;\;\;\;
\overline{{\bf 5}}_{16_H} \;\;\;\;\;\;\;\;\; {\bf 5}_{10_H}
\;\;\;\;\;\;\;\;\;
\overline{{\bf 5}}_{16^{\prime \prime}_H} \;\;\;\;\;\;\;\;\;
{\bf 5}_{\overline{16}^{\prime \prime}_H} \\ \\
\left[ \begin{array}{c} \overline{2} \\ \overline{3} \end{array} \right]
\begin{array}{c} \leftrightarrow \\ \leftrightarrow \end{array}
\left[ \begin{array}{c} 2 \\ 3 \end{array} \right]
\begin{array}{c} \leftrightarrow \\ \leftrightarrow \end{array}
\left[ \begin{array}{c} \overline{2} \\ \overline{3} \end{array} \right]
\begin{array}{c} {\rm ~} \\ \leftrightarrow \end{array}
\left[ \begin{array}{c} 2 \\ 3 \end{array} \right]
\begin{array}{c} \leftrightarrow \\ \leftrightarrow \end{array}
\left[ \begin{array}{c} \overline{2} \\ \overline{3} \end{array} \right]
\begin{array}{c} \leftrightarrow \\ \leftrightarrow \end{array}
\left[ \begin{array}{c} 2 \\ 3 \end{array} \right]
\begin{array}{c} \leftrightarrow \\ {\rm } \end{array}
\left[ \begin{array}{c} \overline{2} \\ \overline{3} \end{array} \right]
\begin{array}{c} \leftrightarrow \\ \leftrightarrow \end{array}
\left[ \begin{array}{c} 2 \\ 3 \end{array} \right] \\ \\
\end{array}
\end{equation}

\noindent
One sees from this that the doublets form two connected, isolated sets.
One of these sets has three doublets, the other five doublets.
In the set with three doublets, two are $\overline{{\bf 2}}$'s
and one is a ${\bf 2}$. Consequently, there is
left massless exactly one $\overline{{\bf 2}}$, which is a linear combination
of the those in ${\bf 16}'_H$ and ${\bf 10}_H$. Similarly, in the isolated
set of five doublets, there are three ${\bf 2}$'s and two
$\overline{{\bf 2}}$'s. So there is left massless exactly one ${\bf 2}$,
which is a linear combination of those in $\overline{{\bf 16}}'_H$,
$\overline{{\bf 16}}^{\prime \prime}_H$, and ${\bf 10}_H$.
The triplets also can be seen to form two connected, isolated sets.
One of these sets contains six triplets and the other contains two.
As each set has an equal number of ${\bf 3}$'s and $\overline{{\bf 3}}$'s,
all the triplets acquire GUT-scale masses. Now, suppose that the only Higgs
doublets that couple to the quarks and leptons are in ${\bf 16}'_H$ and
$\overline{{\bf 16}}^{\prime \prime}_H$. The triplets in these
multiplets can be seen from Eq. (8) to lie in different sets, so that they
are not connected directly or indirectly with each other. Thus the
diagram in Fig. 1 vanishes and the color-triplet-Higgsino-mediated proton
decay is completely suppressed. The doublets in these multiplets are part of
the linear combinations that give the two massless doublets that are the
doublets of the MSSM. 

There is an interesting way of inducing a nonzero $\mu$ term of order the
SUSY breaking scale in this class of $SO(10)$ models.  To see this
mechanism, note that the $(1,1,0)$ component (under standard model gauge
group) of ${\bf 16}_H'$, call it $\nu'^c_H$, will acquire a VEV of 
order $M_{\rm SUSY}$,
once supersymmetry breaking trilinear $A$ terms are included in the Higgs
potential.  There is a tadpole $M_{\rm SUSY} M_{\rm GUT}^2 \nu'^c_H$ that
arises from the first term of Eq. (6).  The scalar component of $\nu'^c_H$
is superheavy with 
a mass-squared of order $M_{\rm GUT}^2$, which upon minimization will
lead to $\left\langle \nu'^c_H\right\rangle \sim M_{\rm SUSY}$. Now,
The $F$ term associated with ${\bf 16}'_H$ contains the term
$\overline{\bf 16}_H({\bf 45}_H+{\bf 1}_H)^*(\overline{\bf 16}'_H{\bf 45}_H)$
(from the first and third terms of Eq. (6)).  Given that  $\left\langle 
\nu'^c_H\right\rangle \sim M_{\rm SUSY}$, this term will induce a VEV of
order $M_{\rm SUSY}$ along the $B-L$ singlet direction of ${\bf 45}_H$.
This then, will induce a $\mu$ term of order $M_{\rm SUSY}$ for the light
Higgs doublets of MSSM from the third term of Eq. (6).  

In sum, we have found a way to do doublet-triplet splitting in $SO(10)$
in a technically natural fashion, with only one adjoint, just two light
Higgs doublets, and complete suppression of the dangerous Higgsino-mediated
proton-decay amplitude. That is, all four of our criteria are fulfilled.
We will now present another, slightly different example.

\vspace{0.2cm}

\noindent
{\it (iii) A second nearly minimal model.}

Consider the following Higgs superpotential terms:

\begin{equation}
\begin{array}{ccl}
W_{2/3} & = & \overline{{\bf 16}}_H ({\bf 45}_H + {\bf 1}_H) {\bf 16}'_H
+ \overline{{\bf 16}}'_H ({\bf 45}_H + {\bf 1}_H) {\bf 16}_H \\
& & \\
& & + \overline{{\bf 16}}^{\prime \prime}_H {\bf 45}_H {\bf 16}'_H +
m \overline{{\bf 16}}^{\prime \prime}_H {\bf 16}^{\prime \prime}_H \\
& & \\
& & + {\bf 16}^{\prime \prime}_H {\bf 16}_H {\bf 10}_H +
\overline{{\bf 16}}_H \overline{{\bf 16}}_H {\bf 10}_H \\
& & \\
& & + {\bf 16}_H {\bf 16}_H {\bf 45}_H {\bf 10}_H/M.
\end{array}
\end{equation}

\noindent
A comparison with Eq. (6) shows a great similarity. There are the same
fields; the first two terms are again the same ``minimal coupling" of
adjoint to spinors as in Eq. (5); and there are the same number and types
of terms. The last term in Eq. (9) is the same kind of quartic coupling
as the last term in Eq. (6), except that here the two spinor fields are
identical. This means that the term automatically gives a 2-mass,
irrespective of how it arises. It could even arise from Planck-scale physics.
One does not, therefore, have to assume that there is some vector that is
integrated out to produce it. To that extent, this is somewhat more
economical than the previous model.

As in the previous model, one can write down a four-by-four mass matrix for
the doublet and triplet Higgs:

\begin{equation}
(\overline{{\bf 5}}_{16_H}, \overline{{\bf 5}}_{16'_H},
\overline{{\bf 5}}_{16^{\prime \prime}_H}, \overline{{\bf 5}}_{10_H})
\left( \begin{array}{cccc}
0 & \langle(45_H + 1_H)\rangle & 0 &  \langle 45_H 16_H \rangle_2 \\ \\
\langle(45_H + 1_H)\rangle & 0 &  \langle 45_H \rangle_3 & 0 \\ \\
0 & 0 & m & \langle 16_H \rangle \\ \\
\langle \overline{16}_H \rangle & 0 & 0 & 0 \\ \end{array} \right)
\left( \begin{array}{c}
{\bf 5}_{\overline{16}_H} \\ \\ {\bf 5}_{\overline{16}'_H} \\ \\
{\bf 5}_{\overline{16}^{\prime \prime}_H} \\ \\ {\bf 5}_{10_H} \\
\end{array} \right).
\end{equation}

\noindent
And, as before, the implications of this mass matrix are most easily
analyzed if one displays the masses in the diagrammatic way used in
Eq. (8):

\begin{equation}
\begin{array}{l}
\;\; \overline{{\bf 5}}_{10_H} \;\;\;\;\;\;\;\;\; {\bf 5}_{\overline{16}_H}
\;\;\;\;\;\;\;\;\; \overline{{\bf 5}}_{16'_H} \;\;\;\;\;\;\;\;\;
{\bf 5}_{\overline{16}^{\prime \prime}_H} \;\;\;\;\;\;\;\;\;
\overline{{\bf 5}}_{16^{\prime \prime}_H} \;\;\;\;\;\;\;\;\; {\bf 5}_{10_H}
\;\;\;\;\;\;\;\;\;
\overline{{\bf 5}}_{16_H} \;\;\;\;\;\;\;\;\;
{\bf 5}_{\overline{16}'_H} \\ \\
\left[ \begin{array}{c} \overline{2} \\ \overline{3} \end{array} \right]
\begin{array}{c} \leftrightarrow \\ \leftrightarrow \end{array}
\left[ \begin{array}{c} 2 \\ 3 \end{array} \right]
\begin{array}{c} \leftrightarrow \\ \leftrightarrow \end{array}
\left[ \begin{array}{c} \overline{2} \\ \overline{3} \end{array} \right]
\begin{array}{c} {\rm } \\ \leftrightarrow \end{array}
\left[ \begin{array}{c} 2 \\ 3 \end{array} \right]
\begin{array}{c} \leftrightarrow \\ \leftrightarrow \end{array}
\left[ \begin{array}{c} \overline{2} \\ \overline{3} \end{array} \right]
\begin{array}{c} \leftrightarrow \\ \leftrightarrow \end{array}
\left[ \begin{array}{c} 2 \\ 3 \end{array} \right]
\begin{array}{c} \leftrightarrow \\ {\rm } \end{array}
\left[ \begin{array}{c} \overline{2} \\ \overline{3} \end{array} \right]
\begin{array}{c} \leftrightarrow \\ \leftrightarrow \end{array}
\left[ \begin{array}{c} 2 \\ 3 \end{array} \right] \\ \\
\end{array}
\end{equation}

One sees immediately that there are exactly two light doublets:
a $\overline{{\bf 2}}$ ($= H_d$), which is a mixture of the doublets
in ${\bf 16}'_H$ and ${\bf 10}_H$, and a ${\bf 2}$ ($= H_u$), which is
a mixture of the doublets in $\overline{{\bf 16}}'_H$,
$\overline{{\bf 16}}^{\prime \prime}_H$, and ${\bf 10}_H$.
The triplets, which all get GUT-scale mass, fall into two sets that
are not connected to each other. If, then, one imagines that only
${\bf 16}'_H$ and $\overline{{\bf 16}}'_H$ couple to quarks and leptons,
the triplets in them will not be connected by a mass term and the
dangerous proton-decay amplitude will be completely suppressed.
On the other hand, each of these multiplets will partially contain one
of the light doublets.  In this model, 
top quark mass arises
through the terms $(16_i 16_j){\bf \tilde{10}}_H + \overline{{\bf 16}}_H
\overline{{\bf 16}}'_H {\bf \tilde{10}}_H$ after integrating out the
Higgs field ${\bf \tilde{10}}_H$.  
 
Another, less economical,
model of doublet-triplet splitting that meets all four criteria
is presented in Appendix A.

We have seen that the doublets of the MSSM can be contained in
spinors as well as vectors of $SO(10)$. An interesting question is whether
a model can be constructed satisfying the four criteria we have stated
that has no vector Higgs at all, but only one adjoint Higgs and
spinor Higgs. The answer is shown to be no in Appendix B. One requires
adjoint, spinor, and vector Higgs. Otherwise there are unacceptable
goldstone fields.

\vspace{0.2cm}

\noindent
{\bf (D) Quark and lepton masses with $\langle {\bf 45}_H \rangle \propto
I_{3R}$}
\vspace{0.2cm}

One thing that any theory of grand unified symmetry breaking must be
able to do is account for, or at least accommodate, the known facts
about quark and lepton masses. As is well known, the minimal $SU(5)$
model gives the unrealistic relations $m_{\mu}^0/m_{\tau}^0 =
m_s^0/m_b^0$ and $m_e^0/m_{\tau}^0 = m_d^0/m_b^0$ (the superscript zero
referring here to masses evaluated at the GUT scale), whereas empirically
one has the Georgi-Jarlskog relations $\frac{1}{3} m_{\mu}^0/m_{\tau}^0 =
m_s^0/m_b^0$ and $3 m_e^0/m_{\tau}^0 = m_d^0/m_b^0$ \cite{gj}.
Various interesting group-theoretical explanations of these Georgi-Jarlskog
factors of 3 have been proposed in the literature \cite{gjth}. Some of
these depend upon the VEV of some adjoint of $SO(10)$ being proportional to
$B-L$ direction. What makes the $B-L$ generator useful in this regard is that
it gives $1/3$ when acting on quarks and $-1$ when acting on leptons.
Thus, a term of the form ${\bf 16}_i {\bf 16}_j {\bf 10}_H {\bf 45}_H$,
where $\langle {\bf 45}_H \rangle \propto B-L$, gives contributions to
the ij elements of the down quark mass matrix $D_{ij}$ and the charged
lepton mass matrix $L_{ij}$ that are in the ratio -1:3.

At first glance, it would seem that having just a single adjoint whose VEV
is proportional to $I_{3R}$ rather than $B-L$ would make it difficult to
avoid the bad relations $m_{\mu}^0/m_{\tau}^0 =
m_s^0/m_b^0$ and $m_e^0/m_{\tau}^0 = m_d^0/m_b^0$. The apparent problem
is that $I_{3R}$ gives exactly the same result when acting on down quarks
and charged leptons. Moreover, the only other superlarge VEVs available to
us in breaking $SO(10)$ are those of the spinors ${\bf 16}_H$ and
$\overline{{\bf 16}}_H$ (or of ${\bf 126}_H$ and
$\overline{{\bf 126}}_H$), which leave unbroken an $SU(5)$; and as we
noted minimal $SU(5)$ also gives the same masses to down quarks and
charged leptons. However, this problem is only apparent.

The symmetry that is responsible for the bad mass relations in the minimal
$SU(5)$ model is an $SU(4)$ subgroup of $SU(5)$. However, this symmetry
is not respected by the VEV $\langle {\bf 45}_H \rangle \propto I_{3R}$.
Therefore this VEV does allow the bad mass relations to be avoided.

A particularly simple way to do this is the following. Consider the three
types of effective Yukawa terms ${\bf 16}_3 {\bf 16}_3 {\bf 10}_H$,
${\bf 16}_2 {\bf 16}_3 {\bf 10}_H {\bf 45}_H$, and
$[{\bf 16}_2 {\bf 16}'_H]_{10} [{\bf 16}_3 {\bf 16}_H]_{10}$. In the
second term, the adjoint VEV is assumed to be proportional to $I_{3R}$.
In the third term it is assumed that the fields in the parentheses
$[...]_{10}$ are
contracted into an $SO(10)$ vector, that the field ${\bf 16}_H$ has a
GUT-scale VEV, and that the ${\bf 16}'_H$ has a weak-scale
$SU(2)\times U(1)$-breaking VEV. These terms give contributions to the down
quark mass matrix $D$ and the charged lepton mass matrix $L$ that have the
following form (see \cite{abb} for explanation)::

\begin{equation}
L = \left( \begin{array}{ccc} 0 & 0 & 0 \\ 0 & 0 & \epsilon' \\
0 & \rho + \epsilon & 1 \end{array} \right) m, \;\;\;\;
D = \left( \begin{array}{ccc} 0 & 0 & 0 \\ 0 & 0 & \rho + \epsilon' \\
0 & \epsilon & 1 \end{array} \right) m.
\end{equation}

\noindent
The entries denoted by 1 come from the operator
${\bf 16}_3 {\bf 16}_3 {\bf 10}_H$. The entries denoted by $\epsilon$ and
$\epsilon'$ come from ${\bf 16}_2 {\bf 16}_3 {\bf 10}_H {\bf 45}_H$.
As expected these entries are exactly the same for the matrices $D$ and $L$,
since the generator $I_{3R}$ does not distinguish leptons from quarks.
Note the important point that $\epsilon$ and $\epsilon'$ are independent
parameters because there are two distinct $SO(10)$
contractions of the fields in the operator.
The entries denoted $\rho$ come from
$[{\bf 16}_2 {\bf 16}'_H]_{10} [{\bf 16}_3 {\bf 16}_H]_{10}$.
As expected, this operator, which involves the spinor VEV that breaks
$SO(10)$ down to $SU(5)$, also gives the same contribution to $D$ and $L$.
However, the crucial point is that this contribution is transposed
between $D$ and $L$. The reason is the well-known fact that $SU(5)$
relates $D$ to $L^T$. This transposition is all-important as it allows
$m_{\mu}^0/m_{\tau}^0 \neq m_s^0/m_b^0$. In particular,
one sees that for $\epsilon, \epsilon' \ll \rho, 1$ the matrices in Eq. (12)
give $m_{\mu}^0/m_{\tau}^0 \cong \rho \epsilon'/(1 + \rho^2)$ and
$m_s^0/m_b^0 \cong \rho \epsilon/(1 + \rho^2)$. To obtain the Georgi-Jarlskog
relation requires only that $\epsilon'/\epsilon \cong 3$.  This model, of course
preserves the good relation $m_b^0 \cong m_\tau^0$.  

This example proves that there is no group-theoretical obstacle to
achieving realistic quark and lepton mass matrices in $SO(10)$ models in
which the only adjoint VEV is proportional to $I_{3R}$. A realistic
and simple model of $SO(10)$ masses including bimaximal
neutrino mixings that is of this type is outlined in
\cite{lopsided}.

\section{$SO(10)$ with two adjoint fields}

We have already seen in section 2 how in a model with two adjoints
doublet-triplet splitting can naturally be done in such a way that
the dangerous proton decay is completely suppressed and there are the right
number of light doublet Higgs. Here we wish to see if it is possible to
construct a simple Higgs sector for such a two-adjoint model.

As was explained in section 3, Part A(iii), the tricky question is how
the spinor Higgs sector is coupled to the adjoint Higgs sector. It turns out
that the method of \cite{barrraby} can be generalized in a very simple way
to the case of two adjoints. Consider the terms

\begin{equation}
\begin{array}{ccl}
W_{2/3} & \supset & \overline{{\bf 16}}_H (\lambda_1 {\bf 45}_{1H} +
\lambda'_1 {\bf 1}_{1H})
{\bf 16}'_H + \overline{{\bf 16}}'_H (\lambda_3 {\bf 45}_{1H} +
\lambda'_3 {\bf 1}_{3H})
{\bf 16}_H \\ \\
& & \overline{{\bf 16}}_H (\lambda_2 {\bf 45}_{2H} + \lambda'_2
{\bf 1}_{2H})
{\bf 16}^{\prime \prime}_H + \overline{{\bf 16}}^{\prime \prime}_H
(\lambda_4 {\bf 45}_{2H} + \lambda'_4 {\bf 1}_{4H}) {\bf 16}_H,
\end{array}
\end{equation}

\noindent
where $\langle {\bf 45}_{1H} \rangle = a_1 (B-L)$, and $\langle
{\bf 45}_{2H} \rangle  a_2 I_{3R}$. The singlets are assumed to be ``sliding
singlets", whose VEVs take the values which make the $F$ terms of the primed
spinors vanish. Thus, $\lambda_1 \langle {\bf 45}_{1H} \rangle + \lambda_1'
\langle {\bf 1}_{1H} \rangle = \lambda_1 a_1 (B-L - 1)$,
$\lambda_3 \langle {\bf 45}_{1H} \rangle + \lambda'_3
\langle {\bf 1}_{3H} \rangle = \lambda_3 a_1 (B-L -1)$,
$\lambda_2 \langle {\bf 45}_{2H} \rangle + \lambda'_2
\langle {\bf 1}_{2H} \rangle = \lambda_2 a_2 (I_{3R} + \frac{1}{2})$,
and $\lambda_4 \langle {\bf 45}_{2H} \rangle + \lambda'_4
\langle {\bf 1}_{4H} \rangle = \lambda_4 a_2 (I_{3R} + \frac{1}{2})$.

Let us now note which generators of $SO(10)$ are spontaneously broken
by which Higgs multiplets. We will denote the generators by their
Standard Model quantum numbers. The spinors
$\overline{{\bf 16}}_H + {\bf 16}_H$ break $SO(10)$ to $SU(5)$ and thus
break the generators in
$(1,1,+1) + (3,2,\frac{1}{6}) + (\overline{3}, 1, -\frac{2}{3})$ plus
their conjugates (which, of course, form a ${\bf 10} + \overline{{\bf 10}}$
of $SU(5)$), and a generator in $(1,1,0)$. The adjoint whose VEV points
in the $B-L$ direction breaks $SO(10)$ to $SU(3)_c \times SU(2)_L \times
SU(2)_R \times U(1)_{B-L}$, and thus breaks the generators in
$(3, 2, \frac{1}{6}) + (3, 2, -\frac{5}{6}) + (3, 1, \frac{2}{3})$
plus their conjugates. The adjoint that points in the $I_{3R}$ direction
breaks $SO(10)$ down to $SU(4)_c \times SU(2) \times U(1)_{I_{3R}}$, and
thus breaks the generators in $(3, 2, \frac{1}{6}) + (3, 2, -\frac{5}{6})
+ (1,1, +1)$ plus their conjugates.

Let us look at the mass matrix of the multiplets in $(1,1, \pm 1)$, which
we will call $E^{\pm}$. This is a five-by-five mass matrix, which can be
read off from Eq. (12).

\begin{equation}
\left( E^+_{16}, E^+_{16'}, E^+_{16^{\prime \prime}}, E^+_{45_1}, E^+_{45_2}
\right) \left( \begin{array}{ccccc}
0 & 0 & \lambda_4 a_2 & 0 & 0 \\ \\
0 & 0 & 0 & \lambda_1 \langle \overline{16} \rangle & 0 \\ \\
\lambda_2 a_2  & 0 & 0 & 0 & \lambda_2 \langle \overline{16} \rangle \\ \\
0 & \lambda_3 \langle 16 \rangle & 0 & M_1 & 0 \\ \\
0 & 0 & \lambda_4 \langle 16 \rangle & 0 & 0 \end{array} \right)
\left( \begin{array}{c} E^-_{\overline{16}} \\ \\ E^-_{\overline{16}'} \\ \\
E^-_{\overline{16}^{\prime \prime}} \\ \\ E^-_{45_1} \\ \\
E^-_{45_2} \end{array} \right).
\end{equation}

\noindent
There are no entries for the $E^+_{16} E^-_{\overline{16}'}$ or
$E^+_{16'} E^-_{\overline{16}}$ elements since $B-L -1$ vanishes for
$E^+$. The $E^+_{45_1} E^-_{45_1}$ entry, which we have denoted $M_1$,
is non-zero because the $(1,1, \pm 1)$ generators are {\it not} broken by
$\langle {\bf 45}_{1H} \rangle$. One can see from this matrix
that there is exactly one massless $(1,1,+1)$, which is
a linear combination of $E^+_{16}$, $E^+_{45_1}$, and
$E^+_{45_2}$. Similarly, there is exactly one massless $(1,1, -1)$.
These are precisely the linear combinations that get eaten by the Higgs
mechanism.

In the same way, one can write a five-by-five mass matrix for the
multiplets in $(\overline{3}, 1, -\frac{2}{3}) + (3, 1, \frac{2}{3})$,
which we denote by $u^c$ and $\overline{u^c}$.

\begin{equation}
\left( u^c_{16}, u^c_{16'}, u^c_{16^{\prime \prime}}, u^c_{45_1}, u^c_{45_2}
\right) \left( \begin{array}{ccccc}
0 & - \frac{4}{3} \lambda_3 a_1 & 0 & 0 & 0 \\ \\
\frac{4}{3} \lambda_1 a_1 & 0 & 0 & \lambda_1 \langle \overline{16} \rangle
& 0 \\ \\
0 & 0 & 0 & 0 & \lambda_2 \langle \overline{16} \rangle \\ \\
0 & \lambda_3 \langle 16 \rangle & 0 & 0 & 0 \\ \\
0 & 0 & \lambda_4 \langle 16 \rangle & 0 & M_2  \end{array} \right)
\left( \begin{array}{c} \overline{u^c}_{\overline{16}} \\ \\
\overline{u^c}_{\overline{16}'} \\ \\
\overline{u^c}_{\overline{16}^{\prime \prime}} \\ \\
\overline{u^c}_{45_1} \\ \\
\overline{u^c}_{45_2} \end{array} \right).
\end{equation}

\noindent
In this case, the $u^c_{16} \overline{u^c}_{\overline{16}^{\prime \prime}}$
and $u^c_{16^{\prime \prime}} \overline{u^c}_{\overline{16}}$ entries are
zero because $u^c$ has $I_{3R} + \frac{1}{2} = 0$. And the entry
denoted $M_2$ is non-zero because the $(3,1, \frac{2}{3}) + (\overline{3},
1, - \frac{2}{3})$ generators are not broken by $\langle {\bf 45}_{2H}
\rangle$. From this matrix, one sees that there is exactly one massless
$(3, 1, \frac{2}{3})$ state, which is a linear combination of
$u^c_{16}$ and $u^c_{45_1}$, and one massless
$(\overline{3}, 1, -\frac{2}{3})$. These are the linear combinations that get
eaten.

In a similar way, one can check the other kinds of multiplets that exist in
the spinors and adjoints, and verify that there are no goldstone modes
besides those that get eaten by the Higgs mechanism. The terms in Eq. (13)
are thus a satisfactory way to couple the adjoint and spinor Higgs sectors to
each other when there are two adjoints, one pointing in the $B-L$
direction and the other in the $I_{3R}$ direction.

\section{Other groups}

In $SU(5)$ the only natural way to solve the doublet-splitting
problem seems to be the ``missing partner mechanism" \cite{dtsu5}, using the
representations ${\bf 75}_H + {\bf 50}_H + \overline{{\bf 50}}_H$.
The ${\bf 50}_H + \overline{{\bf 50}}_H$ contain ${\bf 3} +
\overline{{\bf 3}}$ that do not have ${\bf 2} + \overline{{\bf 2}}$
partners. These triplets can therefore pair up with the triplets in
${\bf 5}_H + \overline{{\bf 5}}_H$ to give them superheavy masses, without
giving any mass to the doublets in ${\bf 5}_H + \overline{{\bf 5}}_H$.
One then has the situation

\begin{equation}
\left( \begin{array}{c} \overline{2}_{\overline{5}} \\
\overline{3}_{\overline{5}} \end{array} \right)
\begin{array}{c} \\ \longleftrightarrow \end{array}
\left( \begin{array}{c} {\rm other} \\ 3_{50} \end{array} \right) \;\;\;\;\;\;
\left( \begin{array}{c} {\rm other} \\ \overline{3}_{\overline{50}}
\end{array} \right)
\begin{array}{c} \\ \longleftrightarrow \end{array}
\left( \begin{array}{c} 2_5 \\ 3_5 \end{array} \right).
\end{equation}

\noindent
Here ``other" represents all the other components of the
${\bf 50}_H + \overline{{\bf 50}}_H$ besides the triplets. These other
components must be made superheavy. However, the only ways to give mass to
all of these other multiplets is to ``mate", directly or indirectly, those in
${\bf 50}_H$ with their conjugates in $\overline{{\bf 50}}_H$.
This would inescapably have the effect of also mating the ${\bf 3}$ in
${\bf 50}_H$ with the $\overline{{\bf 3}}$ in $\overline{{\bf 50}}_H$.
And that would simply bring back the dangerous proton decay.

If one goes to higher unitary groups, other ways to solve the doublet-triplet
splitting problem are possible. For example, in $SU(N)$ with $N \geq 6$
the sliding singlet mechanism can be implemented as shown in \cite{sliding}.
However, the sliding singlet mechaniusm is in essence just a way
of making the fine-tuned cancellation between the adjoint and singlet VEVs
that one has in ``minimal" $SU(N)$ natural. In our terminology, it gives
rise to ``3-masses", but not to ``2-masses". But we have already argued in
Part B of section 3 that with only 5-masses and 3-masses the dangerous
proton decay cannot be suppressed without ending up with four or more
light doublet Higgs.

In $E_6$ it is certainly possible to have satisfactory models which satisfy
all four of criteria simply by embedding in the obvious way the $SO(10)$
models that we have found. We have looked for other solutions to the
problem that are inherently $E_6$, in the sense that they can be realized in
$E_6$ but not in $SO(10)$, but have found none. There are certain
inherently $E_6$ ways of getting 2-masses and 3-masses, but they seem to
leave uneaten goldstone bosons.

By going to groups that involve more than one factor, such as
the Pati-Salam group $SU(4)_c \times SU(2)_L \times SU(2)_R$ \cite{ps}, the
trinification group $SU(3)_c \times SU(3)_L \times SU(3)_R$ \cite{trinify}, or
flipped $SU(5) \times U(1)$ \cite{flipped}, it is easy to solve the
doublet-triplet
splitting problem, suppress the dangerous proton decay, and have the
right number of light Higgs doublets. However, one typically loses thereby the
unification of gauge couplings, which was our third criterion. (However,
in trinification, one can restore the unification of couplings by imposing
a permutation symmetry on the three $SU(3)$ groups.)

\section{Concluding remarks}

SUSY GUTs have a lot of promising features.  
They can elegantly explain some of the puzzling features of the
standard model, such as the observed structure of fermion multiplets,
their mass and mixing pattern including those of the neutrinos, and the
three disparate strengths of the strong, weak and electromagnetic
interactions.  These models however,
run into a roadblock in rapid proton decay mediated by the colored Higgsinos,
which typically has a rate several orders of magnitude larger than the current
experimental limit.  Rather than viewing the nonobservation of proton decay as a
hint against SUSY GUTs, here we have analyzed ways of suppressing these dangerous $d=5$
proton decay operators naturally and completely while preserving all the
good features of SUSY GUTs.
We set forth four criteria to be satisfied in our search,
based on naturalness and simplicity
-- no fine-tuning, no dangerous proton decay,
unification of gauge couplings, and no more than one adjoint.
We have found that simple models do exist in $SO(10)$ GUTs and
its obvious extensions that satisfy all four criteria.
Moreover, it seems necessary to have the vacuum expectation value
of the sole adjoint field point in the $I_{3R}$ direction, rather
than the traditional $B-L$ direction.  We have shown
that such models can be constructed with an almost minimal Higgs
content, and that realistic quark and lepton masses can result in simple
ways.  Our results would seem to suggest that 
proton decay modes mediated by the gauge boson such as
$p \rightarrow e^+ \pi^0$, deserve more attention, since their rates
are unsuppressed in
our models, which can be estimated to be of order $10^{36}$ yr.  Experimental
efforts to reach this level of sensitivity will be highly desirable.

\section*{Acknowledgments}

KSB wishes to thank the Theory Group at Bartol Research Institute
for the warm hospitality extended to him during a summer visit when
this work started.  The work of KSB is supported in part by DOE
Grant \# DE-FG03-98ER-41076, a grant from the Research Corporation
and by DOE Grant \# DE-FG02-01ER-45684.  The work of SMB is supported
in part by DOE Grant \# DE-FG02-91ER-40626.

\section*{Appendix A}

\noindent{\bf Further example for strong suppression of $d=5$ proton
decay operators}
\vspace*{0.1in}

Here we present a model of doublet-triplet splitting in $SO(10)$ that satisfies
the four criteria laid out in the Introduction, but which is very different
from the models given in Section 3. It is also much less economical.

Consider the following couplings

\begin{equation}
\begin{array}{cl}
W_{2/3}  \supset & m_1 \overline{{\bf 16}}'_{1H} {\bf 16}'_{1H}
+ m_2 \overline{{\bf 16}}'_{2H} {\bf 16}'_{2H}
+ m_3 \overline{{\bf 16}}'_{3H} {\bf 16}'_{3H}
+ m_4 \overline{{\bf 16}}'_{4H} {\bf 16}'_{4H} \\ \\
& + \overline{{\bf 16}}'_{1H} \langle {\bf 45}_H \rangle {\bf 16}'_{2H}
+ \overline{{\bf 16}}'_{4H} \langle {\bf 45}_H \rangle {\bf 16}'_{3H}
+ {\bf 10}_H \langle {\bf 45}_H \rangle^2 {\bf 10}_H \\ \\
& + {\bf 16}'_{1H} \langle {\bf 16}_H \rangle {\bf 10}'_H
+ \overline{{\bf 16}}'_{2H} \langle \overline{{\bf 16}}_H \rangle {\bf 10}_H
+ \overline{{\bf 16}}'_{3H} \langle \overline{{\bf 16}}_H \rangle {\bf 10}'_H
+ {\bf 16}'_{4H} \langle {\bf 16}_H \rangle {\bf 10}_H
\end{array}
\end{equation}

The adjoint has VEV in the $I_{3R}$ direction. Note that there are
altogether five pairs of spinor Higgs and two vectors involved in these terms.
In accordance with the notation used in the text, only the unprimed
spinor Higgs have GUT-scale VEVs. These terms give a six-by-six mass
matrix for the doublet and triplet Higgs.

\begin{equation}
\left[ \overline{{\bf 5}}_{16'_1}, \overline{{\bf 5}}_{16'_2},
\overline{{\bf 5}}_{10},
\overline{{\bf 5}}_{10'},
\overline{{\bf 5}}_{16'_3}, \overline{{\bf 5}}_{16'_4} \right]
\left[ \begin{array}{cccccc} m_1 & 0 & c & 0 & 0 & 0 \\
A_3 & m_2 & 0 & 0 & 0 & 0 \\ 0 & \overline{c} & 0 & (A^2)_2 & 0 & 0 \\
0 & 0 & 0 & 0 & \overline{c} & 0 \\ 0 & 0 & 0 & 0 & m_3 & A_3 \\
0 & 0 & 0 & c & 0 & m_4 \end{array} \right] \left[
\begin{array}{c} {\bf 5}_{\overline{16}'_1} \\ {\bf 5}_{\overline{16}'_2} \\
{\bf 5}_{10'} \\ {\bf 5}_{10} \\
{\bf 5}_{\overline{16}'_3} \\ {\bf 5}_{\overline{16}'_4} \end{array} \right]
\end{equation}

\noindent
Here $c \equiv \langle {\bf 16}_H \rangle$, $\overline{c} \equiv
\langle \overline{{\bf 16}}_H \rangle$, and $A \equiv \langle
{\bf 45}_H \rangle$. The subscript `3' on $A$ in two of the entries
indicates that those entries are ``3-masses", while the subscript `2'
on the $A^2$ indicates that it is a ``2-mass".

The mass matrix of the triplets is obtained by setting $(A^2)_2 = 0$ in
the matrix shown above.
It is apparent that the triplets decompose into
two groups that do not mix with each other. Group I consists of
$\overline{{\bf 5}}_{16'_1}$, $\overline{{\bf 5}}_{16'_2}$,
$\overline{{\bf 5}}_{10}$, ${\bf 5}_{\overline{16}'_1}$,
${\bf 5}_{\overline{16}'_2}$, and ${\bf 5}_{10'}$. Group II consists
of $\overline{{\bf 5}}_{10'}$,
$\overline{{\bf 5}}_{16'_3}$, $\overline{{\bf 5}}_{16'_4}$,
${\bf 5}_{10}$,
${\bf 5}_{\overline{16}'_3}$, and ${\bf 5}_{\overline{16}'_4}$.

The mass matrix of the doublets is obtained by setting $A_3 = 0$.
Let us also, for the moment set $(A^2)_2 = 0$. Then we see that
the doublets also fall into two groups. However, the three-by-three
mass matrix of Group I is of rank =2, as is also the three-by-three
mass matrix of Group II. Thus, there would be a massless ${\bf 2}$
and a massless $\overline{{\bf 2}}$ in each group. However, putting
back $(A^2)_2 \neq 0$, one sees that the only a ${\bf 2}$ from Group I
and a $\overline{{\bf 2}}$ from Group II remain massless. These are the
two doublets of the MSSM. Since their triplet partners are in
different groups that do not mix with each other, the dangerous
proton-decay amplitude is completely suppressed.

\section*{Appendix B}

\noindent{\bf  The need for vector Higgs representation
for complete suppression of $d=5$ proton decay operators}
\vspace*{0.1in}

In order to break $SO(10)$ to the Standard Model one needs both adjoint
Higgs and spinor Higgs fields. It is also clear that the breaking of
the weak interactions can be done by doublets that are contained in
spinors. It is therefore an interesting question whether one can have a
realistic $SO(10)$ model that does not contain any vector Higgs fields
at all. The answer is no. One
cannot satisfy all four criteria laid out in the introduction
without vector Higgs fields.

The proof is simple. Suppose there were spinor Higgs fields and an adjoint
Higgs pointing in the $I_{3R}$ direction, but no vectors. Of the three
ways to obtain 2-masses with such an adjoint that are discussed in
Part C(i) of Section 3, only one does not involve vector Higgs fields,
namely terms of the type
$\overline{{\bf 16}}_H (({\bf 45}_H)^2/M + {\bf 1}_H) {\bf 16}'_H$
and $\overline{{\bf 16}}'_H(({\bf 45}_H)^2/M + {\bf 1}_H) {\bf 16}_H$.  
There is also only one way to get 3-masses, namely terms of the
type $\overline{{\bf 16}}'_H {\bf 45}_H {\bf 16}'_H$.

The important observation is that both of these types of terms treat
the fields in $(3, 2, \frac{1}{6})$ and $(\overline{3}, 2, - \frac{1}{6})$
the same way they treat fields in $(1,2, -\frac{1}{2})$ and
$(1,2, \frac{1}{2})$. The 2-mass terms give masses to
all fields except those having $I_{3R} = \pm \frac{1}{2}$ (see Part
C(i) of Section 3). Thus they give mass to both
$(3, 2, \frac{1}{6}) + (\overline{3}, 2, - \frac{1}{6})$
and to $(1,2, -\frac{1}{2}) + (1,2, \frac{1}{2})$. The 3-mass term
only gives mass to terms with $I_{3R} \neq 0$, and so does not give mass
to any of the fields in
$(3, 2, \frac{1}{6}) + (\overline{3}, 2, - \frac{1}{6})$
or $(1,2, -\frac{1}{2}) + (1,2, \frac{1}{2})$.

Now, the model must have a pair of massless doublets
in $(1,2, -\frac{1}{2}) + (1,2, \frac{1}{2})$; and these must be in
the spinor Higgs, since no components of the adjoint have those
quantum numbers. It follows, then, that the spinors will also contain
exactly one $(3, 2, \frac{1}{6}) + (\overline{3}, 2, - \frac{1}{6})$
of fields that do not get mass from spinor-spinor mass terms.
There is also one $(3, 2, \frac{1}{6}) + (\overline{3}, 2, - \frac{1}{6})$
of fields in the adjoint that do not get mass from adjoint-adjoint
mass terms (because these generators are broken by the VEV of the adjoint).
The massless $(3, 2, \frac{1}{6}) + (\overline{3}, 2, - \frac{1}{6})$
in the spinors may not get mass with those in the adjoint, for then there
would be {\it no}
massless $(3, 2, \frac{1}{6}) + (\overline{3}, 2, - \frac{1}{6})$
fields left at all, which is impossible, for there must be such
goldstones to get eaten by the Higgs mechanism. Consequently, there
are massless $(3, 2, \frac{1}{6}) + (\overline{3}, 2, - \frac{1}{6})$
in the spinors as well as in the adjoint. Half of these get eaten, leaving
uneaten exactly one $(3, 2, \frac{1}{6}) + (\overline{3}, 2, - \frac{1}{6})$.
This would be a disaster for the unification of couplings.

Thus, without vector Higgs, one finds that if there are the right number of
light doublets Higgs, there will be uneaten colored goldstone fields.

\newpage

\begin{picture}(360,216)
\thicklines
\put(72,72){\vector(1,1){18}}
\put(90,90){\line(1,1){18}}
\put(72,144){\vector(1,-1){18}}
\put(90,126){\line(1,-1){18}}
\put(108,108){\line(1,0){36}}
\put(180,108){\vector(-1,0){36}}
\put(176,105){$\times$}
\put(180,108){\vector(1,0){36}}
\put(216,108){\line(1,0){36}}
\put(252,108){\line(1,1){18}}
\put(252,108){\line(1,-1){18}}
\put(288,144){\vector(-1,-1){18}}
\put(288,72){\vector(-1,1){18}}
\put(36,140){$Q(u^c)$}
\put(36,68){$Q(\ell^+)$}
\put(296,140){$Q(u^c)$}
\put(296,68){$L(d^c)$}
\put(140,116){$H_{uc}$}
\put(176,116){$M_3$}
\put(204,116){$H_{dc}$}
\put(171,12){{\bf Fig. 1}}
\end{picture}

\vspace{0.2cm}

\noindent
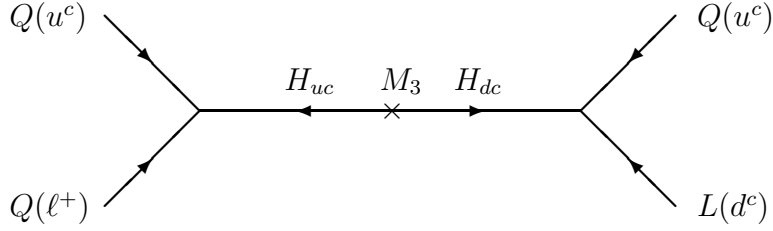
{\bf Figure 1.} The dangerous proton-decay diagram in supersymmetric
grand unified models. $H_{uc}$ and $H_{dc}$ are the color-triplet partners
that the Higgs doublets $H_u$ and $H_d$ have in any unified model.
The fermionic components of $H_{uc}$ and $H_{dc}$ give a dimension-5
proton-decay operator that is generically of order
$M_3/M_{GUT}^2 \sim 1/M_{GUT}$. The $B$ and $L$ violating operators
are of the form $QQQL$ and $\ell^+ u^c u^c d^c$.

\vspace{0.5cm}

\begin{picture}(360,216)
\thicklines
\put(0,72){\vector(1,1){18}}
\put(18,90){\line(1,1){18}}
\put(0,144){\vector(1,-1){18}}
\put(18,126){\line(1,-1){18}}
\put(36,108){\line(1,0){36}}
\put(108,108){\vector(-1,0){36}}
\put(104,105){$\times$}
\put(108,108){\vector(1,0){36}}
\put(144,108){\line(1,0){36}}
\put(176,105){$\times$}
\put(180,108){\line(1,0){36}}
\put(252,108){\vector(-1,0){36}}
\put(248,105){$\times$}
\put(252,108){\vector(1,0){36}}
\put(288,108){\line(1,0){36}}
\put(324,108){\line(1,1){18}}
\put(324,108){\line(1,-1){18}}
\put(360,144){\vector(-1,-1){18}}
\put(360,72){\vector(-1,1){18}}
\put(72,116){$\overline{{\bf 3}}_1$}
\put(136,116){${\bf 3}_2$}
\put(216,116){$\overline{{\bf 3}}_2$}
\put(280,116){${\bf 3}_1$}
\put(100,92){$\langle A \rangle$}
\put(174,92){$M$}
\put(244,92){$\langle A \rangle$}
\put(171,12){{\bf Fig. 2}}
\end{picture}

\vspace{0.2cm}

\noindent
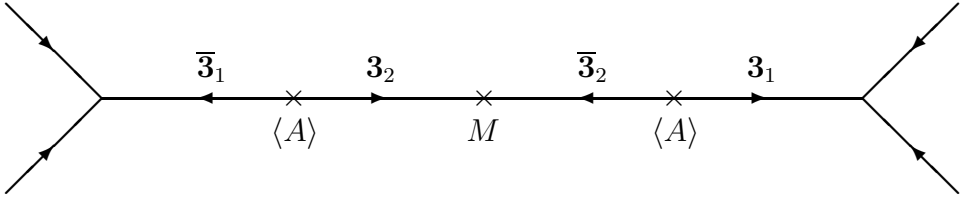
{\bf Figure 2.} The proton decay diagram that results from the couplings
shown in Eq. (3).

\newpage

\begin{picture}(360,216)
\thicklines
\put(72,72){\vector(1,1){18}}
\put(90,90){\line(1,1){18}}
\put(72,144){\vector(1,-1){18}}
\put(90,126){\line(1,-1){18}}
\put(108,108){\line(1,0){36}}
\put(180,108){\vector(-1,0){36}}
\put(176,105){$\times$}
\put(180,108){\vector(1,0){36}}
\put(216,108){\line(1,0){36}}
\put(252,108){\line(1,1){18}}
\put(252,108){\line(1,-1){18}}
\put(288,144){\vector(-1,-1){18}}
\put(288,72){\vector(-1,1){18}}
\put(36,140){${\bf 16}'_H$}
\put(36,68){${\bf 16}_H$}
\put(296,140){${\bf 10}_H$}
\put(296,68){${\bf 45}_H$}
\put(140,116){${\bf 10}'$}
\put(204,116){${\bf 10}'$}
\put(171,12){{\bf Fig. 3}}
\end{picture}

\vspace{0.2cm}

\noindent
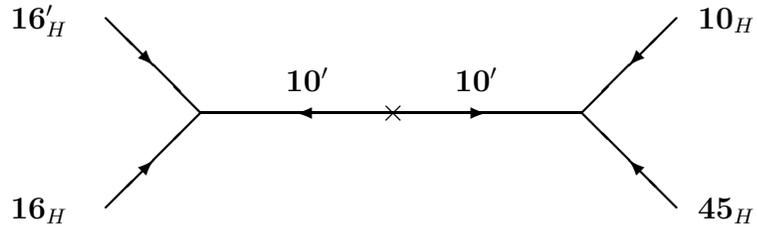
{\bf Figure 3.} A diagram that gives rise to an operator of form
$({\bf 16}'_H {\bf 10}_H) \langle {\bf 16}_H \rangle \langle {\bf 45}_H
\rangle$, which gives a ``2-mass" to the Higgs(ino) fields in
${\bf 16}'_H$ and ${\bf 10}_H$.

\end{document}